\documentclass[preprint,amsmath,amssymb,aps,pre]{revtex4-1}

\usepackage{graphicx}% Include figure files
\usepackage{dcolumn}% Align table columns on decimal point
\usepackage{bm}% bold math
\usepackage{mathrsfs}
\newcolumntype{P}[1]{>{\centering\arraybackslash}p{#1}}
\usepackage{hyperref}% add hypertext capabilities
\begin{document}
\title{Predictive orientational phase behavior in convex polyhedral entropic crystals}
\thanks{mcsad@iacs.res.in}%

\author{Sumitava Kundu}
\author{Kaustav Chakraborty}%
\author{Avisek Das$^\ast$}
\affiliation{School of Chemical Sciences, Indian Association for the Cultivation of Science, Kolkata, INDIA }%

\begin{abstract}
Hard convex polyhedra, idealized models for anisotropic colloids and nanoparticles, are known to form variety of orientational phases despite the regular arrangement of particles in the crystalline assemblies. Based on the orientational behavior of the constituents particles, such phases could be categorized into freely rotating plastic crystals (PC), discrete plastic crystals (DPC) and orientationally ordered crystals (OC). In this article, we report an extensive Monte Carlo computer simulation study of sixty hard convex polyhedral shape indicating a direct predictive relationship between the nature of orientational phases in the crystalline assemblies and single-particle shape attributes. The influence of three attributes namely; (i) Isoperimetric Quotient (IQ) i.e., the extent of asphericity; (ii) isotropy of the moment of inertia tensor in the principal frame and (iii) number of symmetry operations in the point group of the particle and self-assembled crystal structure, were observed to control the orientational phase behavior of the entire solid region in many-body system. The translational order in the crystal appeared to play significant role only in the DPC phase, where as, other two phases were completely governed by the combination of two  attributes. In this study, the role of shape attributes were characterized by sequential appearance of one or two of the aforementioned rotational phases across the phase diagram in a pressure dependent manner which could be regarded as an important stepping stone towards fully predictive self-assembly behavior of hard particle systems.
\end{abstract}

\pacs{Valid PACS appear here}% PACS, the Physics and Astronomy

\maketitle

\section{Introduction}
Crystalline assemblies of colloidal \cite{Pawel1983, Dinsmore1998, Li2016, Orr2022, Boles2016} and nanoparticles (NPs) \cite{Mirkin1996, Robert1996, Whitesides2002, Maye2007,Henzie2012, Bodnarchuk2011, Boneschanscher2014,  Haixin2017, Zhou2022, Geuchies2016, Shevchenko2006, Zhang2013, Jones2010, Glotzer2007} serve the purposes of studying a wide variety of simple and complex behavior of materials. Positional ordering of colloidal crystal and nanocrystal superlattices (NCSLs) have been investigated quite thoroughly since the last few decades leading to a good agreement with experimental studies in contemporary research \cite{Pawel1983, Dinsmore1998, Li2016, Boles2016, Talapin2012, Andrey2002, Tang2005, Glotzer2007, Deng2020, Samanta2022}. The rotational motion of anisotropic particles led to the existence of distinct orientational phases in the structures despite translational order maintained by the particles to reflect the crystalline symmetry. Extensive investigations were carried out experimentally to characterize such phases in the presence of positional ordering \cite{Mirkin1996, Dullens2007, OlveraDeLaCruz2016, Orr2022, Boles2016, Ong2017, Meijer2017} and supported by the computer simulations \cite{Agarwal2011a, Torquato2009, Zhou2024Sci, Zhou2024, Gantapara2015a, OlveraDeLaCruz2016, Kundu2024}. These studies opened up the possibilities of detailed investigations to study the coupling effect between the translational and rotational motion of the particles in crystalline matter. Therefore, the rich phase behavior accomplished by such particles was worthwhile for further investigation, which could lead to the possibility of engineering desired properties as envisioned in atomic and molecular materials. 

Increasing demand of designing new materials and studying material properties gave rise the opportunity to investigate any predictive relationship between the assembled states and shape attributes of the building blocks, if existed. In general, it required a lot of attributes to be taken care of, as the self-assembled states appeared to be contingent on the packing fraction or temperature depending on the situations. The investigators used rigid body dynamics to study the combined effects of particle size and shape by varying the aspect ratio of needle-like particles or tuning the polydispersity as a promising approach to tackle the problem \cite{Phillips2012, Agarwal2012}. The prediction model was also explored in terms of minimum void ratios of the particle size and shape and used to measure the interplay between the particles shapes and different properties of the materials for various engineering applications. A study by Agarwal \textit{et al.} \cite{Agarwal2011a} focused on the ``mesophases'' in hard polyhedral systems to connect the packing fraction dependent behavior of the phases with the particle shapes. This observation turned into extremely relevant in the respective domain, followed by an exhaustive computer simulation study of hard anisotropic particles by Damasceno \textit{et al.} \cite{Damasceno2012c} opening up a new avenue to study the complexity of the aggregates strengthening the requirement of further investigations. More than a couple of attributes i.e., the coordination number of dense fluids and crystals, Isoperimetric Quotient (IQ) and moment of inertia parameter ($\mathcal{M}$) of the particle shape appeared to be pertinent in the respective studies as reported in the literature \cite{Miller2010, Agarwal2011a, Damasceno2012c}. Moreover, patchy colloidal particles were attempted to predict the crystal structures by variable box simulations and using evolutionary algorithm which optimizes the Gibbs free energy \cite{Bianchi2011}. A recent study by Price \textit{et al.} \cite{Price2024} reported a predictive modeling of nanocrystal orientations in the superlattices by using the ligand entropy. The existing studies attempted to outline different predictive relationships in order to make a direct connection between the crystal structures and particle shapes which facilitated the modern day material science research upon designing new kind of targeted materials. As the prediction of entire phase behavior of the aggregated structure invited extreme complexity, it was worthwhile important to compartmentalize the whole problem into a few sub-categories in a systemic manner which could be effective to solve the ultimate goal.

In order to realize the complexity of self-assembly process i.e., complete prediction of assembled states from the particle shapes, if it existed, anisotropic particles interacting via hard-core interaction, were considered as one of the successful models because of the rich phase behaviour originated from the entropy driven assembly of particles shape itself. Hard polyhedral particles as one of the promising prototype of colloidal and nanocrystals, drew attention of the scientific community to study several important physical phenomena, since the assembly proposed by Onsager \cite{Onsager1949}. In particular, freezing and melting of solely entropy driven systems provided a well defined route to theoretical and computational research in colloidal crystallizations \cite{Alder1957, Frenkel1999, Glotzer2007, John2008, Miller2010} revealing many fundamentals aspects of assembled behavior in the light of statistical mechanics. Guided by the aggregation of colloidal and NCSLs into diverse orientational phases in crystal structures, hard convex polyhedra had been subjected to exhaustive computer simulation studies over the past few decades in search of interesting assembly behavior \cite{Glotzer2007, Torquato2009, Batten2010, Ni2012, Agarwal2011a,Damasceno2012c, Gantapara2013, Gantapara2015a, Karas2019, Lee2023, Sharma2024, Kundu2024}. Apart from this, purely entropy driven behavior of hard systems posed interesting fundamental questions about the physical origin of such behavior and nontrivial emergent properties in classical systems. 

As anisotropic particles possess both translational and orientational motions, reducing the entire challenge into the predictive nature of orientational behavior could be a well defined direction assuming the nature of the translational order was well known. The consideration of translational order was important to deal with the coupling effect of different motions of the rigid body with a requirement of adequate solution towards the complete solution. To realize the extent of this problem i.e., complete prediction of orientational behavior in the context of known translational order using the model of hard polyhedral particles, it requires a brief discussion of the possible orientational phases realized in the entropy driven assembly. The freely rotating plastic or rotator crystal is a commonly observed orientational phase which have been widely investigated across multiple disciplines including the domain of nanoscience \cite{Mirkin1996, Zhang2013}, colloidal physics \cite{Meijer2017, Dullens2007} or molecular crystals \cite{Timmermans1961, Reynolds1975, Valsakumar1993, Udovenko2008, Udovenko2010, Vdovichenko2015, Beake2017} etc. These observations were also supported by the computer simulations of hard polyhedra characterizing the true nature of freely rotating motion of the particles staying at the lattice sites dictated by the crystal structures \cite{Torquato2009, Agarwal2011a, Damasceno2012c, Ni2012, Gantapara2013, Kundu2024}. Previous investigations also proved the existence of orientationally ordered states in the crystalline assemblies with all the particles being orientationally ordered in one or multiple directions \cite{Damasceno2012c, Torquato2009, Gantapara2015a, Karas2019, Kundu2024}. Nonetheless, the existence of DPC phase was reported very recently in hard polygons \cite{Shen2019} and hard polyhedra \cite{Lee2023, Kundu2024} where the particles were found to exhibit discrete orientational motion within a fixed set of unique orientations present in the systems. In a recent study, the evidence of a few conserved orientational attributes was reported and the DPC phase was characterized in the context of point group symmetry of the polyhedral particle and self-assembled crystal structure \cite{Kundu2024symm}. It was argued that the existence of this phase was a manifestation of an underlying complex symmetry coupled between the particles and respective crystals reminiscent to the conserved ``local rule'' observed in the ice crystals \cite{Bernal1933, Keen2015}. From the investigations, it was quite evident that almost all kinds orientational phases irrespective of the complexity of translational order, could be possible to achieve using the idealized model of hard particles. As the particle shape alone was the source of complex behavior in the assembled state in such systems, designing the particles in a controlled way could open up the possibility of studying various phases in a systematic manner.

In this article, we presented an extensive Monte Carlo simulation study of sixty convex polyhedral systems, covering the entire pressure range from liquids at low pressure to crystalline solids at highest pressure accessible in simulations. We observed diverse kinds of orientational phase behavior in the crystal structures, the characteristics of which were found to be controlled by a few attributes of the particle shape and symmetry. The rigid polyhedral shapes were categorized based on the similar kinds of orientational phases exhibited in the solid regions followed by analyzing those shapes in terms of multiple shape attributes. Our research indicates the single-particle shape attributes are not enough to make a direct connection with all the orientational phases, rather, the translational order in the crystalline state is also required to know in some cases. To the best of our knowledge, our result provides the strongest evidence of fully predictive relationship between the orientational phases in the crystal structure and the shape attributes of constituent particle. Our findings could be regarded as an important stepping stone before the complete prediction of self-assembled states from the particle shapes which could be treated as a guiding principle of material designing, impacting in the domain of material sciences and applied physics.

\section{Model and methods}\label{model_methods}
We used single-component systems of hard convex polyhedral particles to investigate the orientational phases in crystalline assemblies. Total sixty polyhedra under the influence of hard-core interaction were simulated using Hard Particle Monte Carlo (HPMC) simulation implemented in HOOMD-Blue toolkit \cite{Anderson2016a}. The simulations of either 4096 or 2197 particles were carried out to study the phase behavior of all shapes in the entire solid region under the \textit{NPT} ensemble. The convex polyhedra classified under the classes of Platonic, Archimedean, Catalan, Johnson solids, were chosen in such a way, we could cover a wide range of geometry and symmetry of the particles in shape-space. The polyhedral particles were categorized based on the standard notation depending on Platonic solid (P), Archimedean solid (A), Johnson solid (J) and others (O). The chosen shapes also included both space filling and non-space filling varieties. The translation order of the crystalline structures was verified independently by analyzing the unit cells of the crystal structures showing consistency with the published literature. A few standard techniques were used to analyze the orientational phases in crystal which were reported in earlier \cite{Karas2019, Lee2023, Kundu2024}. The detail simulation protocol and a few analyses techniques to study the orientational phases, are discussed briefly in the following subsections. 

\subsection{Simulation protocol}\label{simulation_protocol}
A dilute system was first prepared with particles ($N$ = 4096 or 2197) on a simple cubic lattice, followed by slow compression of the simulation box, after each step of which, constant volume Monte Carlo (MC) simulation was run and the overlaps are removed. This process was continued up to a packing fraction ($\phi$) of $\sim 55\%-60\%$. Then constant volume MC was carried out around 100 million MC steps until the crystal structures were obtained. The reduced pressure ($P^{\ast}$) of the system was calculated using the scaled distribution function method implemented in HOOMD-Blue toolkit at the particular packing fraction. The pressure $P^{\ast} = \beta P v_0$ where $\beta = \dfrac{1}{k_{B}T}$ and $v_{0}$ was particle volume, set to 1.0 to equalize the packing fraction with the volume fraction. The system was further equilibrated at $P^{\ast}$ value under the \textit{NPT} ensemble allowing for full anisotropic fluctuations of the simulation cell. Thereafter, the pressure $P^{\ast}$ was slowly ramped up; at each value, constant pressure MC steps was performed to equilibrate the system properly. This process was continued up to the a high value of pressure beyond which no noticeable change of the system was observed. The entire process typically involved 5-6 stages and a minimum of 50 million MC steps (MCS) was run at each step. The highest possible packing fraction i.e. densest state explicitly depended on the shape. The most compressed state obtained at the highest value of pressure was taken as the starting point for a series of fresh set of constant pressure melting simulations where the pressure was reduced very slowly and at each step, the system was subjected to long MC run ($\sim$ 50 - 100 million MCS) until the system was properly equilibrated. This process resulted in slow melting of the crystal from the densest state and provided well equilibrated realizations of the system at every packing fraction all the way down to the liquid state. The entire melting regime typically consisted of $\sim 25-27$ state points. The production part of the trajectory at each packing fraction was used to perform all analysis. The whole process was repeated multiple times starting from new starting configurations for checking the statistical significance of the results. Some discrepancies were found between multiple copies for some nearly spherical shapes at the highest packing fraction where the Face centered cubic (FCC) and Hexagonal closed packed (HCP) structures occurred at different independent runs. This was an obvious expectation for sphere-like particles as indicated by the literature \cite{Frenkel1984}. Besides that, we encountered no significant differences across multiple copies of the simulations which agreed well with the published literature. Overall, there were no qualitative changes of the translational and orientational order of the assembled states and we had a solid set up to make the final conclusions of our work as discussed in the Section \ref{sec:results}.

\subsection{Translational and orientational analyses of the systems}
The crystalline order of the assembled states were characterized by analyzing the unit cells where the orientational order and disorder was assigned by coupling the particle orientations with the translational order as introduced in our previous study \cite{Kundu2024}. The orientational statistics of the unit cells was collected to conclude the ordered or disordered state of the crystal structure which provided a well defined and straightforward route to characterize the phases. Once translational order was evaluated, the orientations of the particles were brought under scrutiny to identify the states observed in the respective phases diagrams.
Here, we briefly discuss a few standard techniques to analyze the particle orientations which were already reported in the previous investigations \cite{Karas2019, Kundu2024}. The corresponding literature was referred as the references upon introducing the basic idea, when necessary.

\subsubsection{Pairwise orientational differences}\label{sec:pairwise_differences}
The anisotropic particle orientations in three dimensions were described by quaternions \cite{Anderson2016a}. The orientations of $i$th and $j$th particles were represented by $\mathcal{Q}_i$ and $\mathcal{Q}_j$ respectively and orientational difference between two particles could be estimated by implementing the Equations \ref{eq:equation1},\ref{eq:equation2} consecutively. The point group of the convex polyhedron was taken care of and defined as $\mathcal{Q}^p_{\gamma}$, where $\gamma = 1,2,\ldots,n_p$ and $n_p$ is the number of rotational operations in the corresponding point group. The orientational difference $\theta_{ij}$ could be calculated as the minimum of all angles $\theta^{p}_{\gamma}$ obtained from the Equation \ref{eq:equation1}.
\begin{eqnarray}\label{eq:equation1}
\theta^{p}_{\gamma} = 2 \cos^{-1}[\Re(\mathcal{Q}_i^{\dagger} \mathcal{Q}_j \mathcal{Q}^p_{\gamma})] \:\:\textrm{ for }\gamma = 1,2,\ldots,n_p
\end{eqnarray}
\begin{eqnarray}\label{eq:equation2}
\theta_{ij} = \min \{\theta^{p}_1, \theta^{p}_2,\ldots, \theta^{p}_{n_P}\}
\end{eqnarray}

Upon measuring all pairwise orientational differences in the system, histogram of the angles was evaluated indicating the overall nature of the particle orientations of the system. For completely random orientations, the distribution appeared to be ``shark-fin'' in nature (see Fig. \ref{fig:snapshot_pairwise_dists} for details) whereas, the distribution produced a peak at $\sim$ $0^{\circ}$ signifying all particles to have identical orientations barring statistical noise \cite{Karas2019, Kundu2024}. Though such histograms could precisely characterize the orientational behavior of the particles in completely random state or orientationally ordered state with single orientation, these lacked to provide conclusive information about the system in case there existed more than one orientations. Hence, identifying the unique orientations in the system was required to analyze the phase properly.

\subsubsection{Detection of the unique orientations}\label{sec:unique_orientations}
To detect the number of unique orientations in the simulated system, angular tolerance ($\theta_{c}$) was required to be imposed to deal with the noise. In our previous study, we introduced an algorithmic approach to detect the number of unique orientations which did not depend on the system size and required minimum human interventions \cite{Kundu2024}. In principle, two particles are called orientationally identical if the pairwise angles between two particles $\theta_{ij}$ = 0. This idea was implemented for each frame of the simulation trajectory. For a particular frame, one reference particle was chosen from the system followed by the estimation of all pairwise angles $\theta_{ij}$ with other particles. The value of $\theta_{c}$ corresponded to the first minimum value in the distribution of $\theta_{ij}$ for the respective system which clearly justified the criteria of choosing orientationally ordered particles within the statistical noise. The particles were categorized as a disjoint set $\mathbb{O}_1$ in such a way, so that all particles formed $\theta_{ij} \le \theta_{c}$ with the others in the set. Thus, another reference particle was considered outside of the previous set, and same protocol was applied to construct disjoint set following same criteria. This process was continued until all the particles were found in any of the sets $\mathbb{O}_1, \mathbb{O}_2,\ldots,\mathbb{O}_{k}$. Total number of such disjoint sets corresponded to the total number of unique orientations present in the system $k = N_{\Omega}$. This algorithmic approach gave enough justification to discard the sets with very less number of particles compared to the others. We were able to calculate the population densities in all the unique orientations by estimating the particles in each of the disjoint sets.

The ensemble averaged distribution of all pairwise angles $\theta_{ij}$ appeared to measure the orientation-orientation coupling in the entire system. The plastic crystalline phase could be characterized by the distribution profile $\theta_{ij}$ and comparing the same with that of a synthetically prepared random distribution considering the respective point group of the particle. The distribution of $\theta_{ij}$ could be treated as an order parameter to distinguish any other orientational phases from random state. The detection of unique orientations, population densities in the orientations at each frame of the simulation trajectory was necessary to identify the DPC phase as indicated by the previous investigation \cite{Kundu2024}.

Moreover, to characterize a crystalline state as orientationally order or disorder, the unit cell statistics based on the particle orientations was required to be determined as discussed in the previous section \cite{Kundu2024}. Without performing rigorous free energy calculation of such systems which was not trivial though, these analyses tools were enough to distinguish the orientational phases appeared in the crystals of hard particles. The polyhedral shapes were categorized based on the sequential appearance of orientational phases exhibited in the respective phase diagrams. Thus, the polyhedral shapes corresponding to any specific category were considered and varieties of shapes attributes were tested to find any correlation if existed. Though all the orientational phases were not shown explicitly for all shapes, we verified all the phases showing good agreements with the published results and our observations have been categorized in tables as discussed later. In this article, we mainly investigated the role of various shape attributes controlling such phases and studied the relationships between these entities. The details of our observations are presented in the context of predictive relationships in Section \ref{sec:results}.

\subsubsection{Shape attributes used for predicting orientational phases} \label{sec:shape_att}
A rigid convex polyhedron could be described by several shape attributes. In general, the shape attributes could be extended with new type of complex descriptors incorporating geometrical properties of the shape, it was instructive to seek for correlation(s) between shape and assembly behavior in context of commonly used attributes. It could be also possible, that the emergent behavior was completely driven by the effect of many-body system as it was the most obvious contender to frame the solution intuitively. There existed multiple shape features which could play important role in the framework and those were investigated explicitly. Space-filling shapes were obvious candidates for ordered assembly and the space-filling arrangement could be inferred to be the assembled state at infinite pressure. However, previous studies showed that assembly behavior of space-filling shapes at finite but sufficiently high pressure did not exactly correspond to the space-filling structure \cite{Damasceno2012c}. We used another few attributes to study the predictive relationship of these shape featuring the orientational phases in the crystal structures. Anisotropic particles could be expressed in terms of the respective geometries and point group symmetries. The geometry of a particle could be defined by using multiple obvious parameters like the number of faces, edges, vertices, dihedral angles, asphericity etc., where as the point group of the particle was unique and independent of the geometry. Nevertheless, a number of rigid-body properties could be associated to quantify the motion of the anisotropic particles. Among all the shape features, some could have direct influences on the orientational phases or the combination of these attributes might be relevant in determining the existence of the phases in crystalline states. Simultaneously, different scenario was anticipated to be taken care of where the single-particle shape attributes could not solely determine the orientational phases, the assembled structure was also required to be considered to draw the conclusions. Here, we discuss some of the shape attributes which appeared to be relevant in our investigations.

Anisotropic nature of the shape was the primary reason for complexity in the entropy driven assembly, even though nothing further could be said from a rigorous theoretical standpoint. It was natural to use the measure for the extent of anisotropy to understand the assembly as indicated by the previous investigations \cite{Miller2010, Agarwal2011a, Damasceno2012c}. In this studies, one possible measure of ``asphericity'' was used in the form of \textit{Isoperimetric Quotient}, $IQ = 36 \pi V^{2}/S^{3}$, where $V$ and $S$ denote the volume and surface area of a particle respectively. Our shapes covered a range of values in these parameters, representing a varied degree of anisotropy. It was understandable that more spherical shapes would show a propensity to form plastic phases, even though, it was still not clearly justifiable from a theoretical point of view. This observation was made by Damasceno \textsl{et}\,\textsl{al.} \cite{Damasceno2012c}, most of the shapes in the plastic crystal regions were roughly spherical.  

Another important shape attribute originates from the property of rigid body; the moment of inertia (MOI) expressed in the principal frame of a particle as outlined by previous reports \cite{Miller2010, Agarwal2011a}. The diagonal elements of the tensor are comparable between shapes because of the same volume and uniform mass densities of all particles. Any difference in the MOI is reflective of the difference in geometry according to Equation \ref{eq:moi}.
\begin{equation}\label{eq:moi}
\mathcal{M} = 1 - \dfrac{(I^2_{xx}-I^2_{yy})^2 + (I^2_{yy}-I^2_{zz})^2 + (I^2_{zz}-I^2_{xx})^2}{2(I^2_{xx} + I^2_{yy} + I^2_{zz})^2}
\end{equation}
$I_{xx}$, $I_{yy}$, $I_{zz}$ are the moment of inertia values along the principal axis. This attribute measures the anisotropy in the MOI in the principal frame, which does not necessarily depend on the measure of asphericity. The isotropy in MOI values would necessarily mean the value of $\mathcal{M}$ to be one; the anisotropy in the MOI values corresponded to the value less than one.

To measure the shape correlation with various orientational phases, another feature was used which was not solely dependent on the single particle shape, instead, it accounted for the order of crystallographic point group beside the same of the shape itself. Hence, the order of point group meant total number of operations in the corresponding point group. The number of rotational symmetry operations in the point group intuitively gave the impression of rotational behavior of a rigid body. It directly controlled, in isolation, the number of distinct rotational states the particle could have. The order of particle point group was compared with the same of crystallographic point group. The point groups of these two entities i.e., particle and corresponding crystal structure, were considered and the difference between the order of these point groups was defined as $\mathcal{O}$, where $\mathcal{O} = \mathcal{O}^{c} - \mathcal{O}^{p}$ where $\mathcal{O}^{c}$ and $\mathcal{O}^{p}$ being the order of the point groups of the crystal and the particle respectively. By construction, $\mathcal{O}$ always contains an integer value. This formulation suggests, if a single component system of any polyhedral particles with a certain point group order crystallizes into a structure having the point group of same order, the value of $\mathcal{O}$ equals to zero. Any mismatch between the order directs the value of $\mathcal{O}$ to be either positive or negative integer.

These attributes were considered to correlate the orientational phases under the circumstances of known translational order of the system. We categorized the shapes based on the appearance of particular orientational phase in the corresponding phase diagram and tried to decipher the affinity of the shape features, if it could exist. The detail observation of the relationship are reported in the Section \ref{sec:results}. These observations were completely data-driven and we were unable to justify exact theoretical arguments. This correspondence was tested over a large data set and it appeared to be satisfied by all the shapes reported in this study under the influence of hard-core interaction. We like to point out that the predictive relationship presented here is empirical and does not rule out other situations where such relationship could materialize.

\section{Results}\label{sec:results}
In this study, we explored the orientational phases of self-assembled crystalline solids of sixty polyhedral shapes depending on the packing fractions. All the shapes used in this study, produced crystals as reported in the previous study \cite{Damasceno2012c}. We studied the detail phase behavior of these shapes where a few shapes exhibited solid-solid positional transitions between two crystal structures. In such cases, the changes in orientational order coincided with the positional transitions. The rest of the shapes showed purely orientational transitions with no changes in the positional ordering. The data of two polyhedral shapes namely, Rhombic Dodecahedron (RD) and Elongated Pentagonal Dipyramid (EPD) have been reported to demonstrate different types of orientational phases in the respective crystal structures. The detail study of the phase behavior of these two shapes were already examined previously \cite{Karas2019, Kundu2024}. The positional and orientational phases of all shapes are outlined as the table format in the Section \ref{sec:tables}. Also, we were unable to characterize the orientational behavior of the particles with \textit{Icosahedral} ($I_{h}$) point group at very high packing fractions. These shapes maintained FCC positional ordering in the entire solid region and exhibited PC phases at lower range of packing fractions before the transition to isotropic liquid phase. In the high pressure solids, a few anisotropic choices of the particle orientations could be preferable. Due to extremely highly spherical nature of these shapes, the exact characterization of the orientational phases remained obscured and we had to discard these shapes to maintain the scientific clarity. Such phases could be analyzed in the light of free energy calculation and the development needs further attention which is under strong consideration and will be discussed in a future publication.

\begin{figure*}
	\centering
	\includegraphics[width=1.0\linewidth]{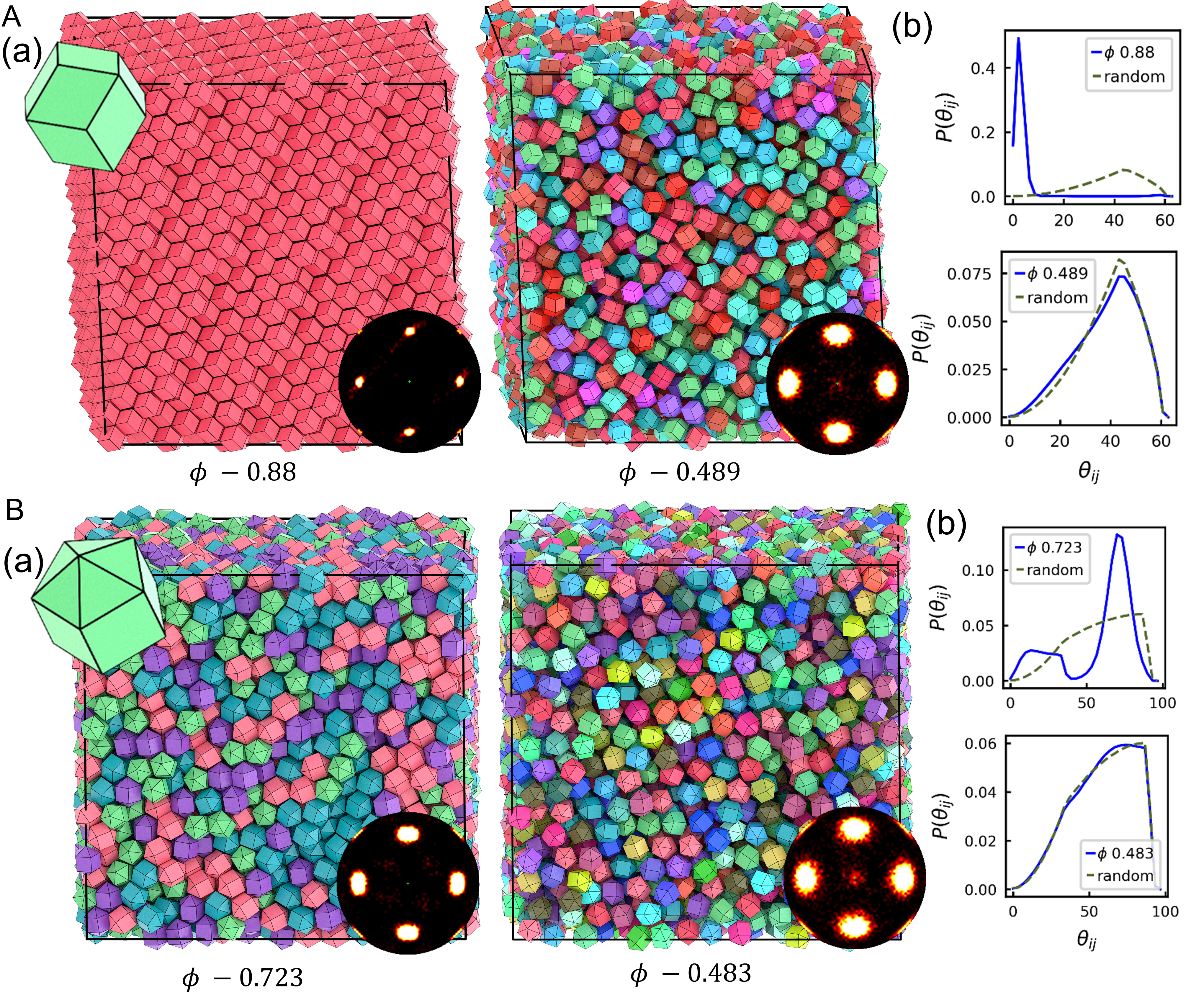}
	\caption{\textbf{Simulation snapshots and different orientational phases are shown at different packing fractions for two shapes:} \textbf{(A)} Rhombic Dodecahedron (RD) shape and \textbf{(B)} Elongated Pentagonal Dipyramid (EPD). The distributions of all pairwise angles are displayed at two packing fractions; one at the densest state and the other at a state just before the melting to isotropic phase. The snapshots are colored according to the unique orientations of the particles. The particle shapes and bond orientational order of crystals are shown as insets for the respective snapshots. The RD system shows OC phase at $\phi$ $\sim$ 0.88 and PC phase at $\phi$ $\sim$ 0.489. The PC phase is compared with synthetically prepared completely random orientational state with good matching of the corresponding distribution. The EPD system exhibits DPC phase at $\phi$ $\sim$ 0.723 with four unique orientations in the system and PC at $\phi$ $\sim$ 0.483. The three kinds of orientational phases observed in the entropy driven assembly are shown for two polyhedral shapes.}
	\label{fig:snapshot_pairwise_dists}
\end{figure*}

\subsection{Different types of orientational phases in crystals}\label{orien_phases}
The orientational behavior of the polyhedral particles in the crystalline solids could be understood by studying the respective phase diagrams at different packing fractions. At the very high value of pressure, the densest crystalline state could correspond to the OC or DPC phases as reported in the earlier literature \cite{Karas2019, Shen2019, Kundu2024} and exhibit freely rotating PC phases at comparatively lower packing fractions before the systems transitioned into the isotropic phases. The simulation snapshots and the distributions of all pairwise angles shown in Figs. \ref{fig:snapshot_pairwise_dists}A,B for two shapes namely, Rhombic Dodecahedron (RD) and Elongated Pentagonal Dipyramid (EPD) denoting different kinds of orientational phases in respective crystals. These two shapes appeared to form FCC crystals in self-assembly and appeared to maintain the same positional ordering across all packing fractions in the solid regions as reported earlier \cite{Damasceno2012c, Karas2019, Kundu2024}. The distributions of all pairwise angles in the system were measured followed by the detection of unique orientations following the protocol described in Section \ref{sec:unique_orientations}, when necessary. In the snapshots shown in the panel (a) of Figs.\,\ref{fig:snapshot_pairwise_dists}A,B, the particles were colored based on the unique orientations present in the system and the particle shapes are shown in the insets. For each shape, two packing fractions from two extreme ends in the solid part were chosen corresponding to different orientational phases ensuring the fact that there existed no other orientational phases in the respective structures. The RD system at the densest state ($\phi$ $\sim$ 0.88) had the orientationally ordered particles leading to the existence of a single color for all particles in the snapshot (see panel (a) of Fig. \ref{fig:snapshot_pairwise_dists}A). The distribution of $\theta_{ij}$ contained a single peak at $\sim$ $0^{\circ}$ confirming the ordered state in the system which sufficiently deviated from the random state as shown in the top portion of panel (b) of Fig. \ref{fig:snapshot_pairwise_dists}A. The previous investigation of EPD shape showed the existence of DPC phase in the orientationally disordered crystal at high density solid and the existence of discrete hopping within the unique orientations was argued at comparatively lower packing fractions \cite{Kundu2024}. The EPD system had four unique orientations at $\phi$ $\sim$ 0.723 and the orientationally ordered particles are shown by a distinct color in the snapshot (see panel (a) of Fig. \ref{fig:snapshot_pairwise_dists}B). Our previous investigation established the fact that DPC phase consisted of orientationally frozen particles at densest states and discretely mobile particles, hopping within the unique orientations at lower packing fractions maintaining the FCC positional ordering \cite{Kundu2024}. The detail discussion of this phase was not presented in this article as the characterization of DPC phase was carried out previously. It was noteworthy, this phase appeared to be equilibrium in nature and multiple orientational features remained conserved in the entire range of packing fractions where the phase was present \cite{Kundu2024symm}. A specific pattern of the distribution of $\theta_{ij}$ was maintained by the particles in the system. Two distinct peaks with different heights were present in the histogram; one peak arrived within a range from $\sim$ $5^{\circ}$ - $40^{\circ}$ like a plateau containing smaller number of pairs indicating the orientational ordering and the other peak occurred at $\sim$ $72^{\circ}$ containing the number of pairs three times of the other, which corresponded to the orientational disordered pairs (see top portion of panel (b) of Fig. \ref{fig:snapshot_pairwise_dists}B). To analyze the PC phase, distributions of all pairwise angles were measured and these profiles were compared with the synthetically produced ``shark-fin'' distribution as shown by the ``dotted-green'' line in panel (b) of Figs.\,\ref{fig:snapshot_pairwise_dists}A,B. The analyses confirmed that there was no preferred orientations in the plastic phase leading to the existence of particles with multiple colors in the respective snapshots and the distribution of $\theta_{ij}$ coincided with that of random orientations. The plastic phase of RD and EPD systems occurred at $\phi$ $\sim$ 0.489 and 0.483 respectively. Though the ranges of orientational phases were not reported explicitly, overall analyses clearly indicated the differences among the three orientational phases obtained in the entropy driven assembly. The analyzed data for other shapes were not discussed in detail, the same for all the shapes were mentioned in the tables along with different shape attributes (See Section.\,\ref{sec:tables} for details).

\begin{figure*}
	\centering
	\includegraphics[width=0.9\linewidth]{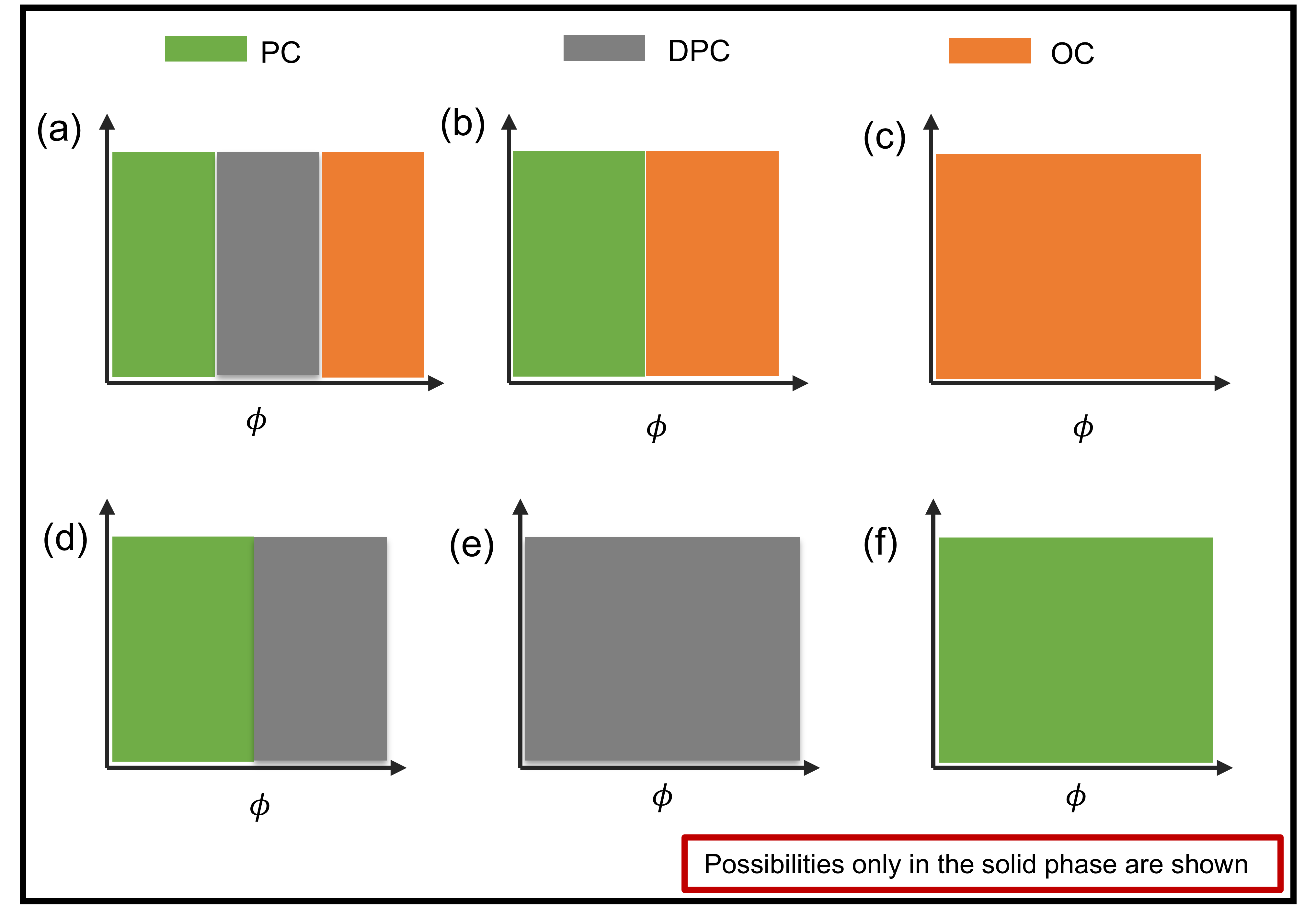}
	\caption{\textbf{Combinations of all possible orientational phases in the crystalline structures are shown in a cartoon representation.} The existence of orientational phases depend on the packing fractions where any single phase or multiple phases can occur in the solid region. The sharp differences between two phases do not carry any significant values of the packing fractions and the packing fraction range of any particular phase corresponds to the representation without any special significance.}
	\label{fig:combination}
\end{figure*}

\subsection{Possible combinations of orientational phases}\label{poss_comb}
Three kinds of orientational phases in the crystalline state could be identified in the entropy driven assembly depending the packing fractions. For a particular shape, multiple orientational phases might appear in the phase diagram. Among the three orientational phases, the PC phase could appear at very low packing fractions before the melting to the isotropic phase where the particles had enough space to exercise the orientational motion. The OC phase was more prone to appear at very high packing fractions in the corresponding phase diagram with an obvious realization that the particles maintained orientational ordering due to unavailability of the free space. As the previous investigation suggested, the DPC phase occurred at comparatively higher range of packing fractions if this phase was indeed present for any shape. This phase was not orientationally ordered but contained a discrete set of unique orientations. The existence of all orientational phases in the crystal structures were realized and the possible combinations have been illustrated in a cartoon representation shown in Fig.\,\ref{fig:combination}. This did not correspond to the fact that all the combinations were possible to achieve for the hard shapes in the respective phase diagrams. It was convincible to observe the appearance of only one kind of orientational phase in the entire solid region. Only the orientationally ordered phase was reported for multiple shapes present as investigated earlier \cite{Damasceno2012c, Damasceno2012b}. For these shapes, it could be a possible explanation that the translational and orientational ordering were strongly coupled and the transition of any specific degrees of freedom affected the other one. Though any direct theoretical argument was not deciphered, it could be observed that in such scenario the crystal structure directly melted to isotropic liquid without exhibiting any further intermediate phase. One another combination, the orientational order and plastic crystal was observed for multiple hard shapes \cite{Agarwal2011a, Damasceno2012c, Karas2019}. Our recent study showed the existence of DPC phase occurring at higher range of packing fractions followed by the PC phase for a few polyhedral shapes \cite{Kundu2024}. In this study, we report the data where three combinations of were obtained; (i) OC and PC phase (panel (b)), (ii) OC phase only (panel (c)) and (iii) DPC and PC phase (panel (d)). We were unable to find any shape where all three phases were present altogether in the phase diagram or any single phase between the DPC and PC existed solely.

\subsection{Relationship of the shape attributes with orientational phases}
In this article, we measured the direct relationship between the particle shapes and different orientational phases in crystal structures. Multiple obvious attributes like number of faces, vertices, edges, dihedral angles were noticed and their characteristics were explored in order to establish straightforward affinity with the orientational phase behavior. These geometric comparisons were fair enough because of the fact, we worked with regular polyhedra with unit volumes. Even though, we could not rule out existence of some sort of complex relationship, nothing obvious seemed to be true and other commonly used attributes turned out to be more useful as discussed below.

\begin{figure*}
	\centering
	\includegraphics[width=1.0\linewidth]{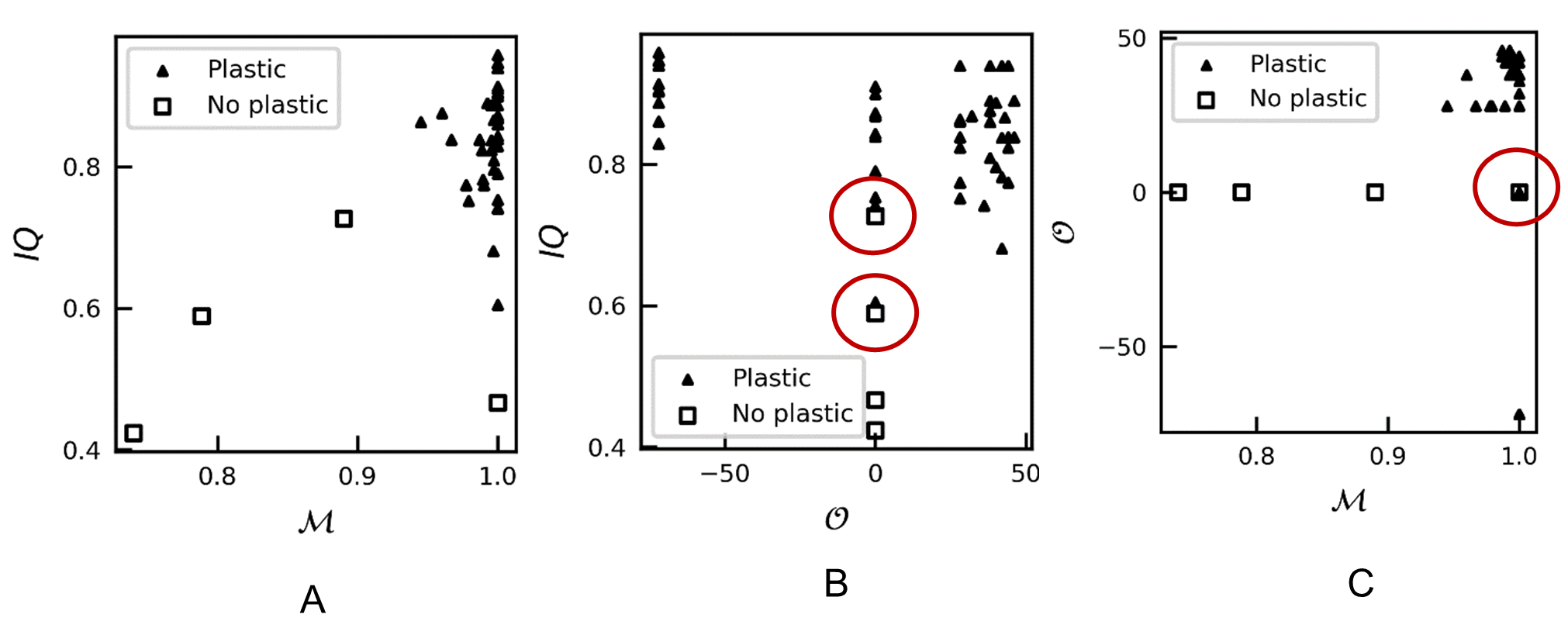}
	\caption{\textbf{The role of particle shapes controlling the plastic crystalline phase are measured in terms of three attributes: } (i) $IQ$, (ii) anisotropy in the moment of inertia values along principal axis ($\mathcal{M}$) and (iii) difference between the order of point groups of the particle and crystal. The data for each attribute are compared with the other to observe the importance of three entities in determining the PC phase. The plot in panel \textbf{(A)} shows the shapes with higher $IQ$ and higher $\mathcal{M}$ values exhibit the plastic phase indicating the important role of these two attributes. The figures shown in panels \textbf{(B)} and \textbf{(C)} suggest no role of the attribute $\mathcal{O}$ to control the plasticity indicating some overlapped area marked by red circles.}
	\label{fig:iq_moi}
\end{figure*}

The role of ``asphericity'' was investigated earlier to study the self-assembled behavior of hard polyhedra \cite{Agarwal2011a, Damasceno2012c, Gantapara2015a}. But, from a more quantitative perspective, incorporating the data across the entire pressure range, only the asphericity clearly did not connect the whole picture. In particular, previous investigation attempted to connect the isotropy of MOI with the mesophases observed in the phase behaviors of a series of convex polyhedra, even though no conclusive outcome was presented \cite{Agarwal2011a}.  We chose these attributes and tried to speculate if any such relationship existed to control the orientational behavior of the particles maintaining the translational order. Our investigation indicated a direct relationship between these shape features and the appearance of plastic crystal phase. We observed that the shapes with high $IQ$ and almost isotropic MOI ($\mathcal{M}$ $\sim$ 1) underscored the existence of PC or rotator phases in the respective phase diagrams. Any deviation from the consideration, i.e., either smaller value of $IQ$ or sufficient anisotropy in the MOI values in the principal frame would correspond to the complete absence of the phase (see Fig.\ref{fig:iq_moi}A). This relationship was valid for all the shapes reported in this study. The $IQ$ and $\mathcal{M}$ values of all the shapes are mentioned in Tables presented in the Section\,\ref{sec:tables}. Any of the two attributes alone was not completely responsible for the existence of plasticity in the solid region. For example, Truncated Tetrahedron (A03) shape, despite having an isotropic MOI ($\mathcal{M}$ = 1), did not show a rotator or PC phase as this shape had a low $IQ$ value (= 0.466). On the other hand, Biaugmented Truncated Cuboctahedron (J67) shape also did not exhibit plastic crystal phase because of the anisotropy in the MOI values along the principal axes despite sufficiently higher value of $IQ$ (= 0.726). From our extensive simulation data and precise analysis, it was clear that both asphericity ($IQ$) and anisotropy in moment of inertial values ($\mathcal{M}$) in the principal frame were important for determining the existence of plastic crystal phase. According to Fig.\,\ref{fig:iq_moi}A, we could loosely define some cutoff values of $IQ$ and $\mathcal{M}$. The shapes with $IQ$ $>$ 0.5 and $\mathcal{M}$ $>$ 0.9, showed the plastic crystalline phases at lower packing fractions. The number of faces and vertices (the number of edges depends on the face and vertices according to Euler's relation) did not explicitly control the existence of plastic phase as shown in the Fig.\,6A. The other attribute $\mathcal{O}$ which depended on the symmetries of both particle and crystal, appeared to be insensitive in determining the existence of plastic crystal phase as shown in Figs.\,\ref{fig:iq_moi}B and C. For multiple shapes, the data of two different phenomena (plastic and non-plastic) almost coincided (marked by the red circles) without any specific conclusion. The orders (number of symmetry elements) of point groups corresponding to all the shapes and self-assembled crystal structures used in the study are given as tables in the Section\,\ref{sec:tables}. Based on the simulation, we could draw the conclusion that there existed no direct affinity of the attributes $\mathcal{O}$, the number of faces, edges, vertices with the appearance of PC phase. It also suggested that the PC was dependent on the single particle shape attribute but not on the crystal structure formed by the shapes. 

\begin{figure}
	\centering
	\includegraphics[width=0.8\linewidth]{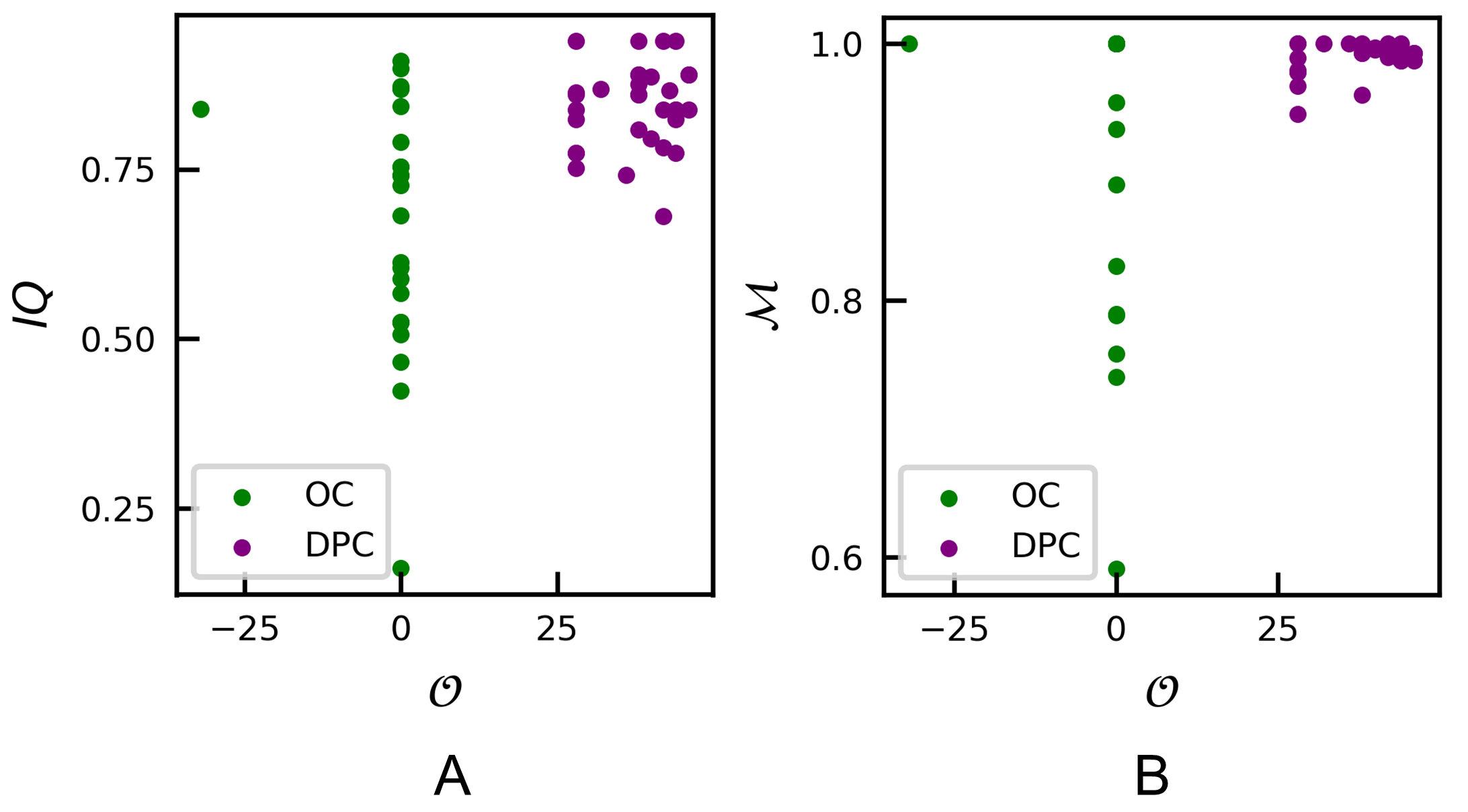}
	\caption{Shapes exhibiting PC phases (triangular markers), are plotted in the space of (A) $IQ$ and $\mathcal{O}$ and (B) $\mathcal{M}$ and $\mathcal{O}$. Plastic forming shapes only with $\mathcal{O}$ $\le$ 0, exhibit ordered phases only, else DPC. $IQ$ and $\mathcal{M}$ do not have any role in determining the existence of the DPC phase at higher pressure range.}
	\label{fig:order}
\end{figure}

Two shape attributes, $IQ$ and $\mathcal{M}$, did not shed any light on other two orientational phases, (i) OC phase and (ii) DPC phase. There were several shapes with lower as well as higher $IQ$ values that gave OC and DPC phases at high pressures  highlighting the fact these parameters were somewhat insensitive to the existence of these phases. We looked at the other attribute $\mathcal{O}$ incorporating the point groups of the particle and crystal structure. The role of particle symmetry was discussed in the context of discrete orientations in the entropy driven assembly \cite{Shen2019, Lee2023}. It turned out that the shapes exhibiting DPC phase at higher packing fractions, had comparatively lower number of symmetry operations in the point group, and it was not related to any way to the other shape features. The lower order of particle point group could not determine the existence of DPC phase alone. As our study revealed, the existence of this phase became more likely for the shapes with $\mathcal{O}$ $>$ 0 as shown in Figs.\,\ref{fig:order}A, B. This meant, in a single component system, if the particle with lower order point group ($\mathcal{O}^{p}$) crystallized into a structure with higher order crystallographic point group ($\mathcal{O}^{c}$) i.e., $\mathcal{O}^{c}$ - $\mathcal{O}^{p}$ $>$ 0, then the DPC phase was expected to appear in the respective phase diagrams. On the other hand, OC phases occurred at higher range of packing fractions for the shapes with $\mathcal{O}^{c}$ - $\mathcal{O}^{p}$ $\leq$ 0. The data presented here underscored the fact, if the particles with higher order point group ($\mathcal{O}^{p}$) formed the crystal structures with equal or comparatively lower order point group then OC phase would more prone to occur. We found the role of point groups of both particle and crystal to control the existence of these two phases at high density solids. The other attributes i.e., number of faces, edges and vertices, $IQ$ and $\mathcal{M}$ were insensitive enough to capture the affinity with these orientational phases (see Figs.\,\ref{fig:order}A, B and Fig.\,6B). It was important to note that the polyhedra with \textit{icosahedral} point group were not considered to study the orientational behavior at high density solids as discussed earlier. The symmetry relationship implemented on all other shapes suggested that there was no straightforward role of single particle shape attributes to establish the correspondence with either the OC or DPC phase in the crystalline structures. The analyses were carried out over the large data set which confirmed the role of these shape features to predict the complete orientational behavior of the particles if corresponding crystal structures were known. 

It was important to realize that, this relationship encountered all the orientational phases separately and the occurrence of any particular phase did not guarantee the appearance of other in the phase diagram. It was evident from this study that we were unable to find the shapes for all possible combinations of orientational behavior as outlined in Fig.\,\ref{fig:combination}. Our analyses indicated the existence of any particular kind of orientational behavior was solely governed by the dedicated attributes. As the study revealed, the orientational behavior at lower packing fractions was controlled by the single-particle shape attributes where as, the behavior at higher range of packing fractions was not determined by the geometry of particle only but the symmetries of both particle and corresponding crystal structure.

\begin{figure}
	\centering
	\includegraphics[width=0.6\linewidth]{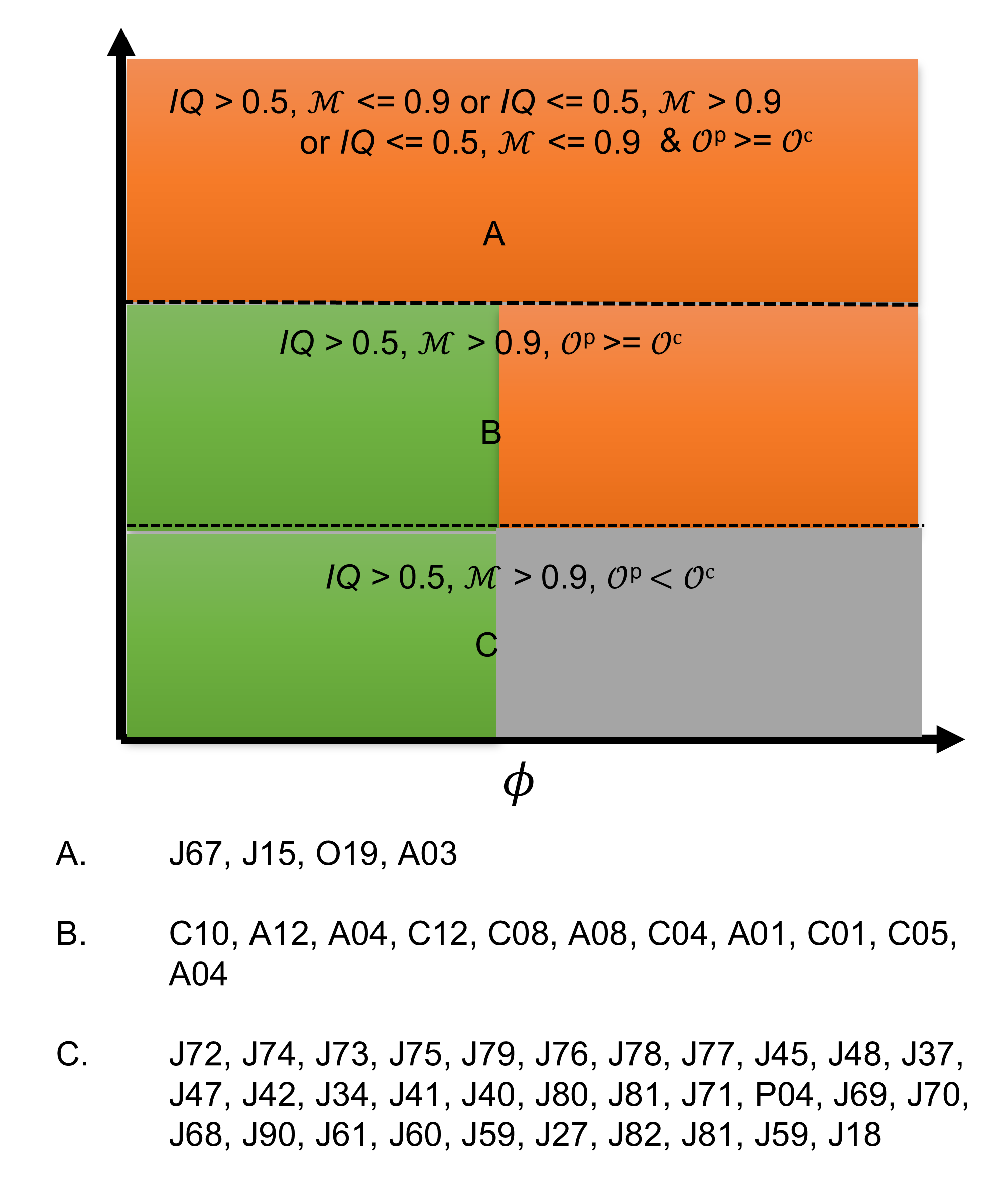}
	\caption{\textbf{Sequential appearance of three combinations of orientational phases are shown in cartoon representation: }\textbf{(A)} Only OC phase, \textbf{(B)} OC and PC phases and \textbf{(C)} DPC and PC phases. The cutoffs for $IQ$ and $\mathcal{M}$ are roughly defined. The shapes satisfying both conditions; $IQ$ $>$ 0.5 and $\mathcal{M}$ $>$ 0.9, exhibit the plastic crystal phases at lower packing fractions. The orientationally ordered phases are present at high packing fractions for the shapes with $\mathcal{O}_{p}$ $\geq$ $\mathcal{O}_{c}$, otherwise the correlated disordered phases will show up. The sharp transition between two phases does not carry any special significance, rather it has been shown for the better visualization only.}
	\label{fig:summary}
\end{figure}

\section{Discussion}\label{sec:discussion}
Combining all the empirical relationships between the shape attributes and orientational phase behavior, we arrived at the following conclusion, which seemed to explain the whole picture in a coherent framework. Three attributes, namely the measure of asphericity (quantified by the $IQ$), anisotropy of the MOI in the principal frame (estimated by $\mathcal{M}$) and difference between the order of point groups of the particle and crystal structure, controlled all orientational phases across the entire solid region. Without a proper theoretical rationale, we cannot rule out the role of other attributes, or a complicated combinations of those. But based on our extensive simulations and analyses, we concluded that a minimalistic model of correspondence between shape and self-assembly in the orientational space of crystalline hard polyhedral systems could be constructed in terms of just three aforementioned attributes. The data of the correlations are given in Fig.\,\ref{fig:iq_moi} and \ref{fig:order}. The relationship is pressure dependent and different attributes appeared to control the behavior of different pressure range of the phase diagram. It was not possible to span the space of these two parameters by the polyhedral shapes (or the exhaustive collections of shapes reported by Damasceno \textsl{et.}\,\textsl{al.}\, \cite{Damasceno2012c}for that matter). The accurate cutoff for the measurement of $IQ$ and $\mathcal{M}$ was hard to assign, but any value equivalent to 0.5 for $IQ$ and 0.9 for $\mathcal{M}$ was stable for acknowledging the orientational phases. The order of any point groups either the particle or crystal, has no role in determining the plastic crystal phase at sufficiently lower pressure range. For shapes with higher  value of $IQ$ compared to the previously defined cutoff, the appearance of the rotator phase was controlled by the isotropy parameter in moment of inertia values i.e. $\mathcal{M}$. For low value of $IQ$ or significant deviation from isotropic moment of inertia (MOI) values along the principal axis ($\mathcal{M}$ $<$ 0.9) also gave the evidence for the absence of plasticity as driven by the extensive data, illustrated by Fig.\,\ref{fig:summary}. These shapes did not have a qualitatively different behavior other than perfect OC or DPC phases, which was somewhat weakened at low packing fraction, but did not show the well defined sign of plasticity and melted into isotropic liquid phase directly. For the shapes that were highly anisotropic with very low IQ, however, MOI values did not have any role and lack of rotator phase persisted. One possible argument could follow the fact, for these shapes the degree of rotation was quite high to adjust the orientations of neighbor particles that would not lead to the possibility of maintaining the same translational order; as a result the crystal structure melted into liquid. Though it was not possible to make any rigorous justification without the theoretical standpoint but the orientational behavior of Truncated Tetrahedron (A03) shape belonged to this category. The region just above the liquid, appeared to be perfectly captured by two dimensional plot with $IQ$ and $\mathcal{M}$ as both the axes, the limiting case was indicated, for which we got at least one test case. The behavior appeared to be continuous in this plane, which is understandable from a physical point of view, however no other insight could be drawn from this empirical relation. We have projected the entire crystal region of the data presented by Damasceno \textit{et al.} \cite{Damasceno2012c} on our predictive analysis parameterized by three attributes. We found that the shapes designated as ``plastic crystal'' perfectly fits to our description. Based on our consideration, we present the data of several shapes that were known to have rotator crystals, also exhibit DPC phases at the high packing fraction as summarized by Fig.\,\ref{fig:summary}. Our investigation could give insights that the existence of DPC phase is not governed by the any individual shape attributes or combinations of those, of a single particle. As this phase occurred at the unit cell level of the crystal, one implication could be drawn based on the order of point group symmetries of both particle and the crystal. From the data set, we also observed another correlation indicating that OC phases with $O_{h}$ point group at the densest packing, always exhibited PC phases at lower pressure regions. Such occurrences might be quite anecdotal which would require rigorous theoretical justification to conclude. Those crystals other than the cubic symmetry at the highest packing density had more possibility to form orientationally ordered structure, either directly melt to isotropic liquid phases without having any signatures of plasticity or exhibit solid-solid positional transitions to have a plastic behavior at lower packing fraction regions. In summary, we provide strong evidence for possible existence of predictive relationship involving just three shape attributes between the shapes and certain feature of the self-assembly, namely the nature of orientational order for regular convex polyhedra. To the best of our knowledge, this is by far the strongest evidence of predictive self-assembly within the subclass of shapes forming crystals as the positionally broken symmetry phase, and takes us one step closer to fully predictive assembly which will involve similar relationship(s) with positional order, ultimately the space group of the assembled crystal.

\section{Conclusions}\label{sec:conclusions}
The data presented in this article could shed light on the coupling effect of translational and rotational motion of the rigid body in the classical many-body system. Different shape features became relevant at different regions of the phase diagram resulting in distinct orientational phases in the crystalline states. Our observation could play important role to streamline the framework of material design in the context of predictive behavior of particles in crystals. Though the complete prediction of the self-assembled state is still unsolved, our investigation could be considered as a stepping stone towards the ultimate solution of the grand problem. 

\begin{acknowledgments}
	We acknowledge financial support from DST-INSPIRE Fellowship (IVR No.\,201800024677) provided to SK. AD thanks DST-SERB Ramanujan Fellowship (SB/S2/RJN-129/2016) and IACS start-up grant. KC thanks IACS for financial support. Computational resources were provided by IACS HPC cluster and partial use of equipment procured under SERB CORE Grant No.\,CRG/2019/006418.
\end{acknowledgments}

% Bibliography
\bibliography{shape_att_hard_poly}

\newpage
\onecolumngrid
\section*{Supplementary Information}
\subsection{Role of vertices, faces, edges in orientational phases} 
\begin{figure}[!h]
	\centering
	\includegraphics[width=0.9\linewidth]{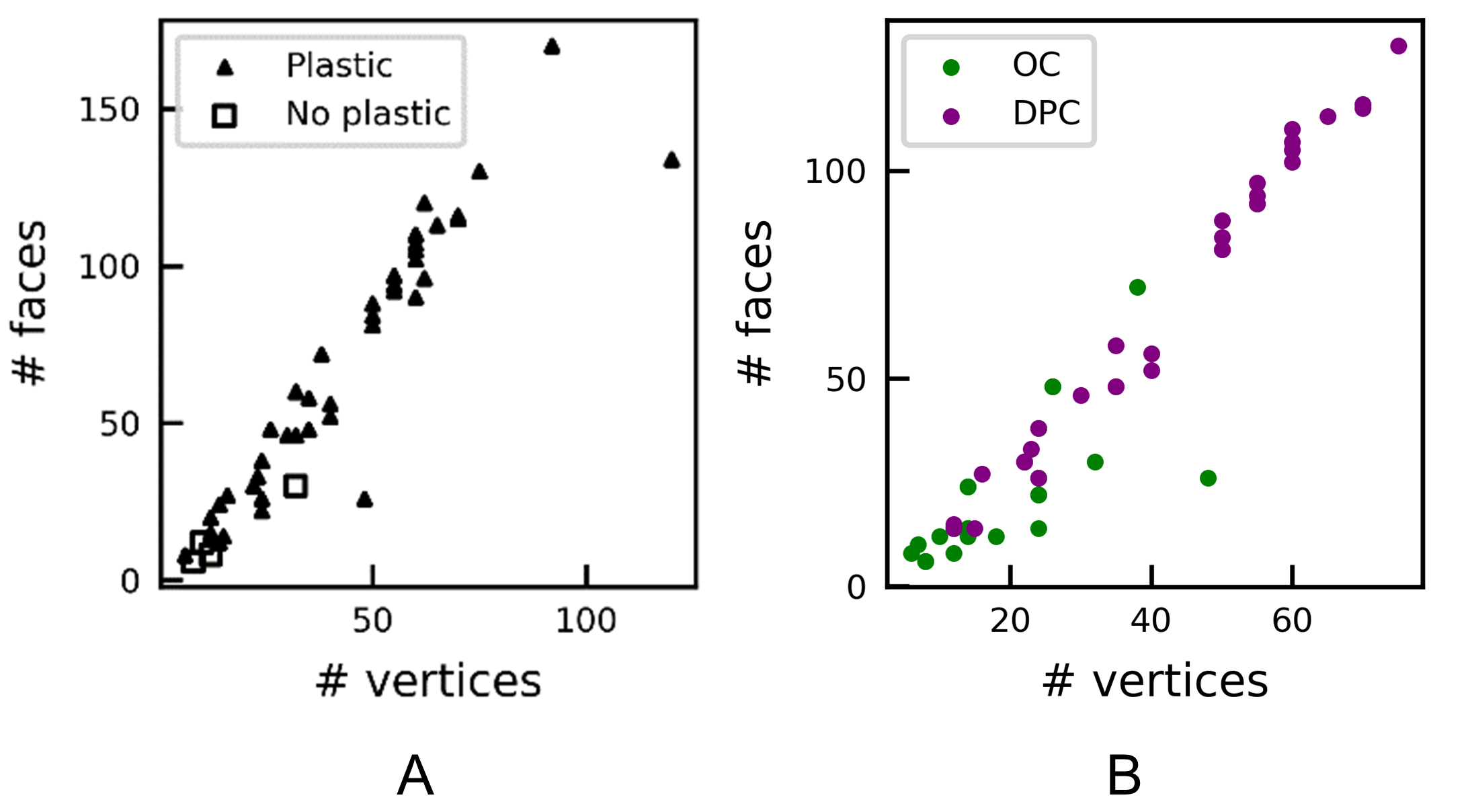}
	\caption{(A) The number of vertices and the number of faces are plotted for the shapes forming the PC and no PC in the phase behavior. (B) The shapes exhibiting OC and PC phases and the shapes having DPC and PC phases are shown in the space of same shape attributes, i.e. number of vertices and the number of faces. Both the plots could not be used to inherit any conclusions to find the predictive relationships.}
	\label{fig:vertices_vs_faces}
\end{figure}

The shapes were plotted in the space of number of vertices and faces in Fig.\,\ref{fig:vertices_vs_faces} to understand the predictive relationship. Fig.\,\ref{fig:vertices_vs_faces}A showed the shapes forming the plastic phases and no plastic phases in the shape space of the number of vertices and faces. The shapes having the OC phases along with the PC phases  and the shapes with DPC and PC phases, are shown in Fig.\,\ref{fig:vertices_vs_faces}B. We were not be able to find any predictive relationships between the orientational phase behavior and the number of vertices and faces. The plots with the number of edges were not shown because the number of edges is dependent on the number of vertices and the faces for a regular convex polyhedron according to the Euler's equation.

\pagebreak
\subsection{Different orientational phases and shape attributes of particles}\label{sec:tables} 
\begin{table}[!h]
	\centering
	{\renewcommand{\arraystretch}{1.7}%
		\begin{tabular}[t]{|P{1cm}|P{2.7cm}|P{2cm}|P{2cm}|P{1cm}|P{1cm}|P{2cm}|P{1cm}|}
			\hline
			Shape & OC & DPC & PC & $IQ$ & $\mathcal{M}$ & $\Gamma^{p}$ & Order\\
			\hline
			C11 & - (FCC) & - (FCC) & Yes (FCC) & 0.95 & 1.0 & $I_{h}$ & 120\\
			\hline
			A13 & - (FCC) & - (FCC) & Yes (FCC) & 0.94 & 1.0 & $I_{h}$ & 120 \\
			\hline
			C13 & - (FCC) & - (FCC) & Yes (FCC) & 0.94 & 1.0 & $I_{h}$ & 120 \\
			\hline
			C09 & - (FCC) & - (FCC) & Yes (FCC) & 0.94 & 1.0 & $I_{h}$ & 120 \\
			\hline
			C06 & - (FCC) & - (FCC) & Yes (FCC) & 0.94 & 1.0 & $I_{h}$ & 120 \\
			\hline
			A09 & - (FCC) & - (FCC) & Yes (FCC) & 0.93 & 1.0 & $I_{h}$ & 120 \\
			\hline
			J72 & No & Yes (FCC) & Yes (FCC) & 0.93 & 1.0 & $C_{5v}$ & 10\\
			\hline
			J74 & No & Yes (FCC) & Yes (FCC)  & 0.93 & 1.0 & $C_{2v}$ & 4 \\
			\hline
			J73 & No & Yes (FCC) & Yes (FCC) & 0.93 & 1.0 &$D_{5d}$ & 20 \\
			\hline
			J75 & No & Yes (FCC) & Yes (FCC) & 0.93 & 1.0 & $C_{3v}$ & 6 \\
			\hline
			A11 & - (FCC) & - (FCC) & Yes (FCC) & 0.91 & 1.0 & $I_{h}$ & 120 \\
			\hline
			C10 & Yes (FCC) & No & Yes (FCC) & 0.91 & 1.0 & $O_{h}$ & 48 \\
			\hline
			C07 & - (FCC) & - (FCC) & Yes (FCC) & 0.90 & 1.0 & $I_{h}$ & 120 \\
			\hline
			A06 & - (FCC) & - (FCC) & Yes (FCC) & 0.90 & 1.0 & $I_{h}$ & 120  \\
			\hline
		\end{tabular}
	\label{table:table1}
	}
\end{table}
\pagebreak
\begin{table}
	\centering
	{\renewcommand{\arraystretch}{1.7}%
		\begin{tabular}[t]{|P{1cm}|P{2.7cm}|P{2cm}|P{2cm}|P{1cm}|P{1cm}|P{2cm}|P{1cm}|}
			\hline
			Shape & OC & DPC & PC & $IQ$ & $\mathcal{M}$ & $\Gamma^{p}$ & Order\\
			\hline
			A12 & Yes (FCC) & No & Yes (FCC) & 0.89 & 1.0 & $O_{h}$ & 48 \\
			\hline
			J79 & No & Yes (FCC) & Yes (FCC) & 0.89 & 0.99 & $C_{1v}$ & 2 \\
			\hline
			J76 & No & Yes (FCC) & Yes (FCC) & 0.89 & 0.99 & $C_{5v}$ & 10 \\
			\hline
			J78 & No & Yes (FCC) & Yes (FCC) & 0.89 & 0.99 & $C_{1v}$ & 2 \\
			\hline
			A04 & Yes (FCC) & No & Yes (FCC) & 0.75	& 1.0 & $O_{h}$ & 48 \\
			\hline
			J77 & No & Yes (FCC) & Yes (FCC) & 0.89 & 0.99 & $C_{5v}$ & 10 \\
			\hline
			C02 & - (FCC) & - (FCC) & Yes (FCC) & 0.88 & 1.0 & $I_{h}$ & 120 \\
			\hline
			J45 & No & Yes (FCC) & Yes (FCC) & 0.88 & 0.99 & $D_{4}$ & 8 \\
			\hline
			J48 & No & Yes (FCC) & Yes (FCC)  & 0.87 & 0.96 & $D_{5}$ & 10\\
			\hline 
			C12 & Yes (FCC) & No & Yes (FCC)  & 0.87 & 1.0 & $O_{h}$ & 48\\
			\hline
			C08	& Yes (FCC) & No & Yes (FCC)  & 0.87 & 1.0 & $O_{h}$ & 48 \\
			\hline
			J37 & No & Yes (FCC) & Yes (FCC) & 0.86	& 1.0 & $D_{4d}$ & 16 \\
			\hline
			A08 & Yes (FCC) & No & Yes (FCC) & 0.86	& 1.0 & $O_{h}$ & 48 \\
			\hline
			J47 & No & Yes (FCC) & Yes (FCC) & 0.86	& 0.99 & $C_{5}$ & 5 \\
			\hline
			J42 & No & Yes (FCC) & Yes (FCC) & 0.86	& 0.94 & $D_{5h}$ & 20\\
			\hline
			A02 & - (FCC) & - (FCC) & Yes (FCC) & 0.86 & 1.0 & $I_{h}$ & 120 \\
			\hline
		\end{tabular}
	\label{table:table2}
	}
\end{table}
\pagebreak
\begin{table}
	\centering
	{\renewcommand{\arraystretch}{1.7}%
		\begin{tabular}[t]{|P{1cm}|P{2.7cm}|P{2cm}|P{2cm}|P{1cm}|P{1cm}|P{2cm}|P{1cm}|}
			\hline
			Shape & OC & DPC & PC & $IQ$ & $\mathcal{M}$ & $\Gamma^{p}$ & Order\\
			\hline
			A02 & - (FCC) & - (FCC) & Yes (FCC) & 0.86 & 1.0 & $I_{h}$ & 120 \\
			\hline
			J34 & No & Yes (FCC) & Yes (FCC) & 0.86 & 1.0 & $D_{5h}$ & 20\\
			\hline
			J41 & No & Yes (FCC) & Yes (FCC) & 0.86 & 1.0 & $C_{5v}$ & 10 \\
			\hline
			J40 & No & Yes (FCC) & Yes (FCC) & 0.86 & 1.0 & $C_{5v}$ & 10 \\
			\hline
			C04 & Yes (FCC) & No & Yes (FCC) & 0.84	& 1.0 & $O_{h}$ & 48 \\
			\hline
			J80 & No & Yes (FCC) & Yes (FCC) & 0.83 & 0.96 & $D_{5d}$ & 20 \\
			\hline
			J81 & No & Yes (FCC) & Yes (FCC) & 0.83	& 0.98 & $C_{2v}$ & 4 \\
			\hline
			J71 & No & Yes (FCC) & Yes (FCC) & 0.837 & 0.996 & $C_{3v}$ & 6 \\
			\hline
			P04 & - (FCC) & - (FCC) & Yes (FCC) & 0.829 & 1.0 & $I_{h}$ & 120 \\ 
			\hline
			J69 & No & Yes (FCC) & Yes (FCC) & 0.824 & 0.989 & $D_{5d}$ & 20 \\
			\hline
			J70 & No & Yes (FCC) & Yes (FCC) & 0.824 & 0.995 & $C_{2v}$ & 4 \\
			\hline
			J68 & No & Yes (FCC) & Yes (FCC) & 0.809 & 0.997 & $C_{5v}$ & 10 \\
			\hline
			J90 & No & Yes (FCC) & Yes (FCC) & 0.795 & 0.997 & $D_{2d}$ & 8 \\
			\hline
			J61 & No & Yes (FCC) & Yes (FCC) & 0.782 & 0.989 & $C_{3v}$ & 6 \\
			\hline
			J60 & No & Yes (FCC) & Yes (FCC) & 0.774 & 0.99 & $C_{2v}$ & 4 \\
			\hline
			J59 & No & Yes (FCC) & Yes (FCC) & 0.774 & 0.978 & $D_{5d}$ & 20 \\
			\hline
		\end{tabular}
	\label{table:table3}
	}
\end{table}
\pagebreak
\begin{table}
	\centering
	{\renewcommand{\arraystretch}{1.7}%
		\begin{tabular}[t]{|P{1cm}|P{2.7cm}|P{2cm}|P{2cm}|P{1cm}|P{1cm}|P{2cm}|P{1cm}|}
			\hline
			Shape & OC & DPC & PC & $IQ$ & $\mathcal{M}$ & $\Gamma^{p}$ & Order\\
			\hline
			J27 & No & Yes (FCC) & Yes (FCC) & 0.741 & 1.0 & $D_{3h}$ & 12 \\
			\hline
			A01 & Yes (FCC) & No & Yes (FCC) & 0.741 & 1.0 & $O_{h}$ & 48 \\
			\hline
			C01 & Yes (FCC) & No & Yes (FCC) & 0.74 & 1.0 & $O_{h}$ & 48 \\
			\hline
			J82 & No & Yes (FCC) & Yes (FCC) & 0.838 & 0.987 & $C_{1v}$ & 2 \\
			\hline
			J81 & No & Yes (FCC) & Yes (FCC) & 0.838 & 0.987 & $C_{2v}$ & 4 \\
			\hline
			C05 & Yes (BCC) & No & Yes (BCC) & 0.79 & 1.0 & $O_{h}$ & 48 \\
			\hline
			A04 & Yes (BCC) & No & Yes (BCC) & 0.753 & 1.0 & $O_{h}$ & 48 \\
			\hline
			J18 & No & Yes (BCC) & Yes (BCC) & 0.681 & 0.997 & $C_{3v}$ & 6 \\
			\hline
			A10 & Yes (BCT) & No & Yes (BCC) & 0.839 & 1.0 & $O_{h}$ & 48 \\
			\hline
			J67 & Yes (BCT) & No & No & 0.726 & 0.89 & $D_{4h}$ & 16 \\
			\hline
			J15 & Yes (BCT) & No & No & 0.589 & 0.789 & $D_{4h}$ & 16 \\
			\hline
			P02 & Yes (BCC) & No & Yes (BCC) & 0.605 & 1.0 & $O_{h}$ & 48 \\
			\hline
			O19 & Yes (Rhombohedral) & No & No & 0.424 & 0.74 & $D_{3d}$ & 12 \\
			\hline
			A03 & Yes (Cubic Diamond) & No & No & 0.466 & 1.0 & $T_{d}$ & 24 \\
			\hline
		\end{tabular}
	}
	\label{table:table4}
\end{table}

\end{document}